# Molecular Characterization of Ebselen Binding Activity to SARS-CoV-2 Main Protease

Cintia A. Menéndez, Fabian Byléhn, Gustavo R. Perez-Lemus, Walter Alvarado and Juan J. de Pablo*.

Pritzker School of Molecular Engineering

University of Chicago

4670, S. Ellis Avenue, Chicago, IL 60637

depablo@uchicago.edu

## Abstract

The Coronavirus Disease (COVID-19) pandemic caused by the SARS-coronavirus 2 (SARS-CoV-2) urgently calls for the design of drugs directed against this new virus. Given its essential role in proteolytic processing, the main protease Mpro has been identified as an attractive candidate for drugs against SARS-CoV-2 and similar coronaviruses. Recent high-throughput screening studies have identified a set of existing, small-molecule drugs that have already been approved for various diseases as potent Mpro inhibitors. Amongst these, Ebselen (2-Phenyl-1,2-benzoselenazol-3-one), a glutathione peroxidase mimetic seleno-organic compound, is particularly attractive. Recent experiments suggest that its effectiveness is higher than that of other molecules that also act at the enzyme's catalytic site. By relying on extensive simulations with all-atom models, in this study we examine at a molecular level the potential of Ebselen to decrease Mpro's catalytic activity. Our results indicate that Ebselen exhibits a distinct affinity for the catalytic site cavity of Mpro. In addition, our molecular models reveal a second, previously unkown binding site for Ebselen in the dimerization region localized between the II and III domains of the protein. A detailed analysis of the free energy of binding indicates that the affinity of Ebselen to this second binding site is in fact significantly larger than that to the catalytic site. A strain analysis indicates that Ebselen bound between the II-III domains exerts a pronounced allosteric effect that regulates catalytic site access through surface loop interactions, and induces a displacement and reconfiguration of water hotspots, including the catalytic water, which could interfere with normal enzymatic function. Taken together, these findings provide a framework for the future design of more potent and specific Mpro inhibitors, based on the Ebselen scaffold, that could lead to new therapeutic strategies for COVID-19.

Introduction:

A new coronavirus of zoonotic origin, SARS-CoV-2, is the etiological agent responsible for the 2019-2020 viral pneumonia COVID-19 outbreak that commenced in Wuhan City, Hubei Province, China (1-4). Currently, targeted therapeutics are not available and, at this point in time, effective treatment options remain tentative and very limited.

Conventional drug design and drug development strategies require years of investigations and substantial investments. The repurposing of approved pharmaceutical drugs and drug candidates already in clinical trials therefore provides an essential, alternative approach to rapidly identify drugs with clinical potential and to manage new infectious diseases that lack specific treatments and vaccines. A recent high-throughput screening study (5) considered more than 10,000 compounds against COVID-19. The study included approved drugs, drug candidates and natural products that were screened by an enzymatic inhibition assay seeking to find inhibitors of the virus's main protease, Mpro; Ebselen, a relatively small molecule, emerged from that work as a new drug lead to target this crucial enzyme. The structures of Ebselen and Mpro are shown in **Figure 1**. Mpro is a key CoV enzyme, which plays a central role in mediating viral replication and transcription, making it an attractive drug target against this virus (6, 7).

Ebselen, which has also shown strong antiviral activity in cell-based assays (5), is an organoselenium compound with anti-inflammatory, anti-oxidative, and cytoprotective properties. It has been investigated in the context of multiple diseases, such as bipolar disorders (8) and hearing loss (9, 10). In addition, a recent report indicated that Ebselen also showed potent bactericidal activity against multidrug-resistant (MDR) clinical isolates of staphylococcus aureus (11). Likewise, the use of silver and Ebselen in synergistic formulations have been shown to be effective against five, clinically difficult-to-treat MDR Gram negative bacteria (12). Ebselen has extremely low cytotoxicity (LD50 in rats > 4,600 mg/kg, per os) (13), and its safety in humans has been evaluated in several clinical trials (9, 10, 14). Taken together, these reports and the underlying data underscore the clinical potential of Ebselen for CoV-2 treatment.

In an effort to understand the molecular mechanisms through which Ebselen interacts with Mpro, in this work we present results of atomistic molecular simulations that provide useful, previously unknown insights into the Mpro-Ebselen complex, and which might provide new avenues to rationally enhance Ebselen's activity. First, we present an analysis of the most probable interaction sites between Mpro-Ebselen, as well as, absolute binding free energy calculation for a complementary quantitative evaluation. Second, we evaluate the role of different binding sites on molecular stiffness and molecular strain. Finally, we examine how Ebselen binding modifies the structure and transport of water in Mpro's catalytic site. We find that Ebselen binds to two sites, as opposed to only one at the catalytic site, and we also identify an allosteric mechanism that influences the catalytic site when Ebselen is bound at the distant site. Our findings are summarized in a concluding section, and several suggestions for future experimental work are presented.

Results:

Molecular dynamics simulations of Mpro and Ebselen were carried out using the AMBER18 simulation package (see Methods section for details). A total of 3 μs of atomistic molecular dynamics trajectories of Mpro using Ebselen as a molecular probe were organized into 15 replicas of 200 ns each. These trajectories were analyzed to construct probability density maps for the preferred locations of Ebselen around the protein. The results are shown in **Figure 1**. Two distinct, highly probable binding sites emerged from this analysis. The first is located within the catalytic site, and the second is in a region that is essential for Mpro's dimerization (15), between the II and III domains. Similar observations were reported in a previous study (16), where simple organic solvents/compounds such as acetonitrile, benzene, dimethylsulfoxide, methanol, phenol, and urea were used as molecular probes. In addition, the authors considered the potential mutability of residues belonging to the Mpro catalytic site, and explained that the development of drug resistance associated with the natural evolution of Mpro could wipe out efforts that target this protein for COVID-19 treatment. Instead, they emphasized the promise of alternative strategies aimed at targeting the region between the II and III domains, which is implicated in dimer formation. The results shown in **Figure 1** are therefore encouraging in that Ebselen appears to target both the catalytic site and the dimerization domain.

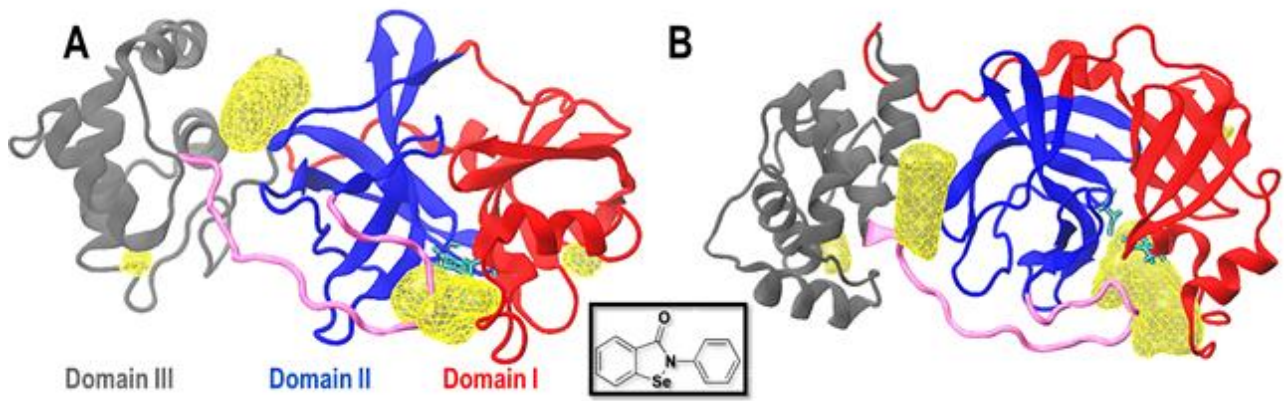

**Figure 1.** Structure of Mpro and density maps of Ebselen binding. **(A)** lateral view, and **(B)** top view. In both figures, yellow surfaces show the most probable interaction sites between Mpro and Ebselen (highest probability density). Mpro domain I, II and III are shown in red, blue and grey, respectively. Cys145 and His41 (the catalytic dyad), are shown in cyan. The loop (residues 185 to 201) connecting domains I and II with domain III is shown in pink. The inset shows the chemical structure of the Ebselen molecule.

**Figure 2** provides a closer look at Ebselen's binding motifs in both regions (the catalytic site and the intersection between domains II–III). In the catalytic site, **Figure 2A**, hydrogen bonding interactions are formed between the carbonyl oxygen of Ebselen and the Asn142 and Gln189 side chains. One can also appreciate the hydrophobic contacts between Ebselen and Met165, Pro168, Met49, His164 and both residues belonging to the catalytic dyad, His41 and Cys145.

**Figure 2B** shows a representative configuration of Ebselen at the intersection between domains II and III. There are distinct hydrophobic contacts with Phe294, Pro108, Ile200, Val202, His246, Thr292, Ile249, Pro132 and Ile249 (some residues are not shown for clarity). Highly dynamic hydrogen bonds were observed between Ebselen and the Gln107, Gln110, and Hie246 side chains.

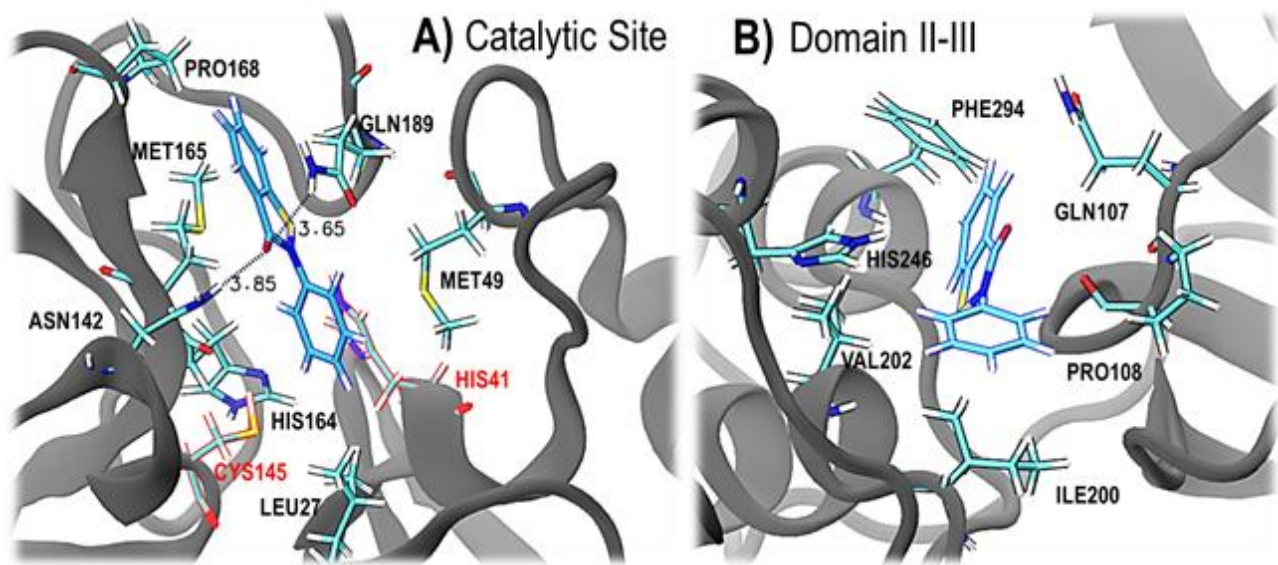

**Figure 2.** Binding modes for Ebselen-Mpro complexes. **(A)** At the catalytic site, and **(B)** at the intersection between Domain II-III. In both figures, Ebselen, as well as main residues displaying contacts, are showed with sticks.

The results shown in **Figure 1** were generated on the basis of direct molecular dynamics simulations. In order to arrive at a quantitative estimate of the binding affinity of each site, we used thermodynamic integration to determine the absolute binding free energy for the catalytic site and the Domain II-II site. More specifically, we used Particle Mesh Ewald Molecular Dynamics

(PMEMD) as implemented in Amber 18 (33) with 11 windows per integration and 10ns per window. In addition, multiple runs starting from the most probably binding cluster identified previous MD simulations were considered. In this way, three independent replicas for each site were used to calculate averages. The results are shown in **Table 1.**

| Binding Site | Absolute Binding Free Energy (kcal/mol) |
|---|---|
| Catalytic Site | -5.55±2.28 |
| Domain II-III Site | -8.87±1.59 |

**Table 1.** Absolute binding free energy for Mpro-Ebselen complex for each binding location. The average and uncertainty energy values were obtained from 3 independent replicas for each site. Details of the calculation are described in Methods.

Consistent with the probability density maps shown in **Figure 1**, the absolute binding free energies corresponding to both sites are negative, serving to underscore the thermodynamic stability of the Ebselen-Mpro complex at both sites. Here we note the binding affinity of Ebselen at the second binding site, located between the II and III domains, is in fact greater that at the catalytic site. A previous report suggested that this small molecule could also inhibit Mpro through non-covalent binding, particularly because it has been found to exhibit a stronger inhibition effect than other compounds that were also able to modify Cys145 in the catalytic dyad quantitatively (5). The results presented here could explain Ebselen's high enzymatic inhibition, even when this compound could only partially modify Cysteine 145.

Having identified a distant binding site for Ebselen between domains II and III, we examined the role of this binding site, if any, on the catalytic site of the protein. To do so, we determined the local strain induced by the binding throughout the protein. To examine potential synergies of drug binding to both sites, we also considered a third scenario in which both sites (catalytic and distant) are simultaneously occupied by Ebselen. Note that local strain provides a measure of local deformation that filters out non-trivial, functional conformational changes from non-functional ones, and it is therefore ideally suited to highlight how a perturbation (i.e. Ebselen binding) at one site induces conformational changes at other, potentially distant, sites. Note that different measures of strain have recently been used in studies of allostery in proteins (17, 18, 19). Here, we used a method originally introduced by our group in the context of local strain in polymeric glasses.

**Figure 3** shows the shear strain throughout the protein upon binding of Ebselen at different sites, alongside the β-factor (estimated from the Root Mean Squared Fluctuation, RMSF), which measures thermal fluctuations for each residue. The shear strains are measured relative to the average conformation from all frames of the apo protein trajectory; such strains, therefore, correspond to the deformation upon binding relative to the apo protein. The shear strain for the apo structure is then showing an average internal strain in the protein.

From the β-factor analysis, it is evident that when Ebselen is bound to the Domain-II-III interface (**Figure 3A**: blue lines), the dynamics of the 44-52 loop, which flank the catalytic site, is significantly altered. On the other hand, when Ebselen is bound at both sites simultaneously there is a clear global reduction of the thermal fluctuations among the receptor except for one single region located around residues 137-140, where enhanced flexibility is apparent. Apart from these residues, Ebselen bound at both sites shows the lowest β-factor values, however, at this specific point, this system exhibits as high flexibility as Mpro-apo protein (**Figure 3A**: green line).

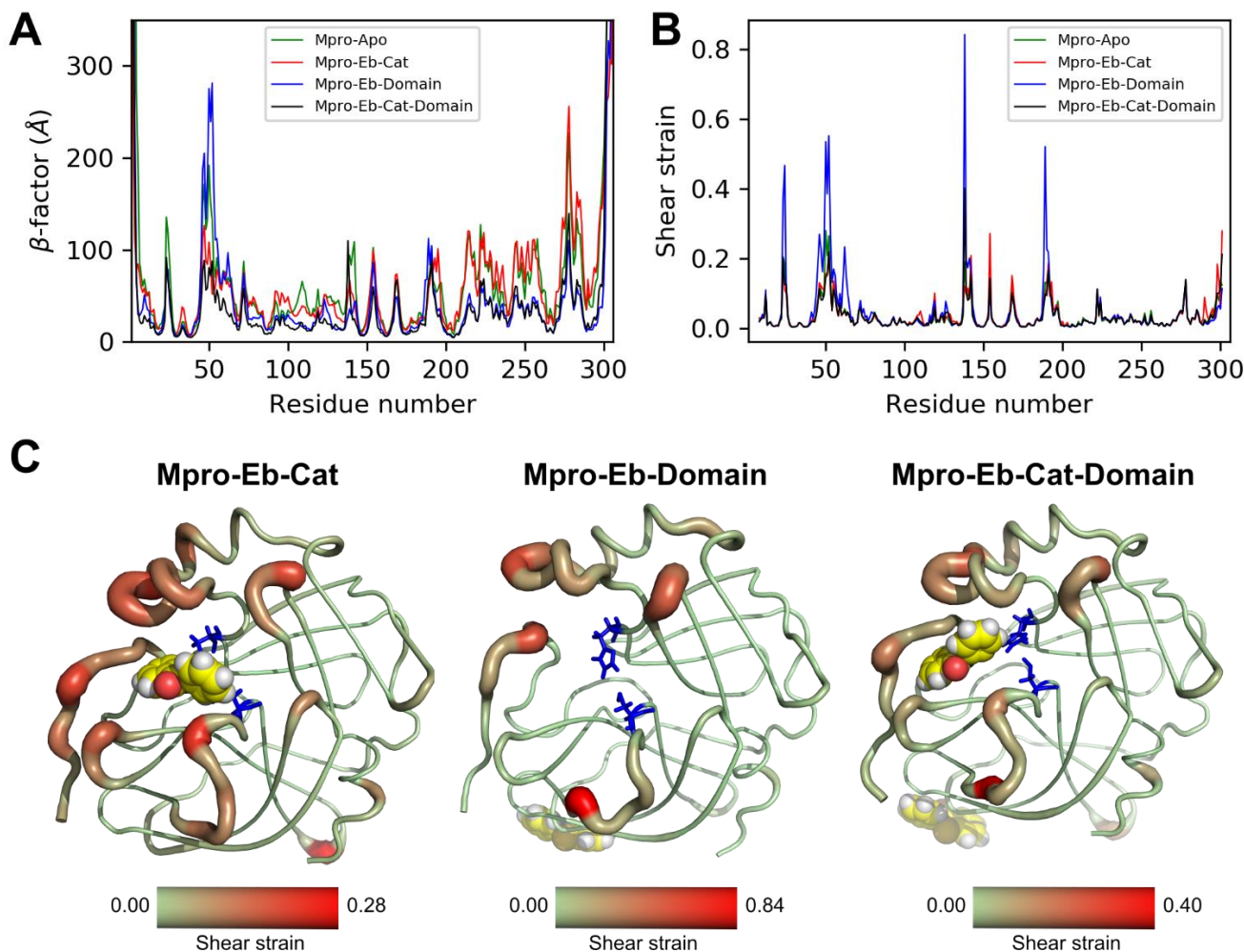

**Figure 3. (A)** β-factor, estimated from the Root Mean Squared Fluctuation, RMSF, **(B)** Shear strains. In both upper panels, Mpro apo results are shown in green (Mpro-apo); for Ebselen, when bound to the catalytic site (Mpro-EB-Cat), results are in red. For Domain II-III interface (Mpro-Eb-Domain), results are shown in blue. For two Ebselen molecules bound at both sites simultaneously (Mpro-Eb-Cat-Domain), results are shown in black. For the shear strain calculation, only the Cα atoms are included. **(C)** Shear strains mapped onto the different protein complexes from (B). The redder the region and the larger the radius of the structure, the higher the shear strain. For clarity, only the catalytic site and domain I-II are shown, the drug is shown as spheres, and the catalytic dyad as blue sticks.

The strain analysis helps disentangle the effects of functional and non-functional fluctuations and provides a more detailed view of the effects of Ebselen binding. As seen in **Figure 3A** for the β-factor analysis, when Ebselen is bound to the domain interface it produces a large strain at the 44-52 loop (**Figure 3B**: blue line). Likewise, the 185-201 loop that also flanks the catalytic site, as well as the region comprising residues 137-140, exhibit a high strain, which is not immediately apparent in the β-factor. When both molecules are bound simultaneously to Mpro (**Figure 3B**, black line), a high strain signal is also exhibited around residues 137-140, as expected based on the aforementioned β-factor results. Finally, when Ebselen is located in the catalytic site, a much lower strain is showed around all these aforementioned three regions (**Figure 3B**, red line). **Figure 3C** shows that the strain is primarily localized at the two loops flanking the catalytic site (44-52 and 185-201 loops), as well as a loop in catalytic site (residues 22-25). A highly strained region also appears around residues 137-140. To understand why the 137-140 residues show such a large strain when Ebselen is bound to both sites we turn our attention to the molecular images shown in **Figure 4**.

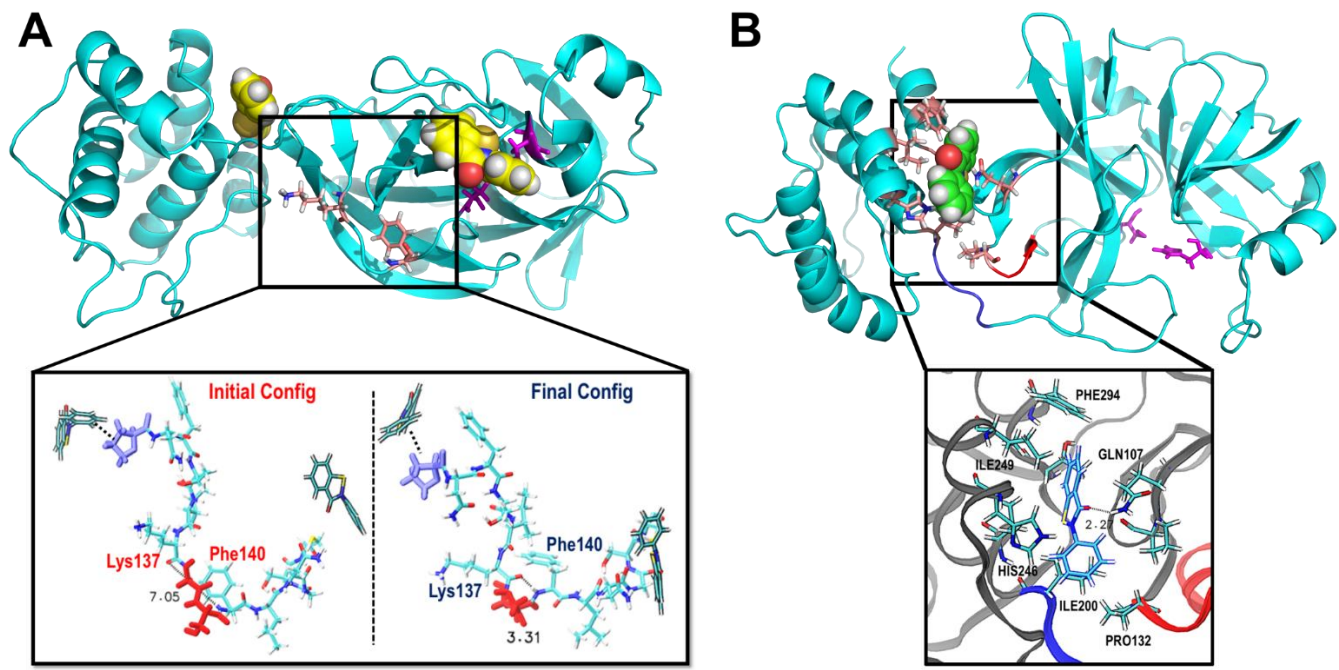

**Figure 4.** Close-up of the highly strained regions identified from strain analysis. **(A)** The Mpro structure with Ebselen molecules bound to both catalytic site and domain II-III interface simultaneously. In the upper panel, Lys137 and Phe140 are shown as sticks. The lower panel shows the close-up of this region and shows the backbone H-bond between Lys137 and Phe140 at the beginning and end of the simulation. **(B)** The Mpro structure with Ebselen bound to domain II-III interface. The close-up shows that a hydrogen bonding interaction is formed with Gln107 side chain, and hydrophobic contacts are formed with His246, Val202, Ile249, Phe294, Pro132, and Ile200. This last residue is located at the end of the 185-201 loop which connects domain III with domain I-II (Catalytic site). In both (A) and (B) upper figures, Ebselen is shown as spheres and the catalytic dyad as magenta sticks.

In the close-up of the residues shown in **Figure 4A**, a specific backbone hydrogen bond is formed between Lys137 and Phe140 due to the binding of Ebselen between domains II and III; this conformational change induced by the presence of Ebselen causes the high strain shown in **Figure 3B**. This conformational change takes place in the middle region between the catalytic site (Cys145) and the binding cleft between domains II and III (Pro132), which points to the relevance of these residues.

From these results, it is evident that when Ebselen binds between the II-III domains, it exerts a pronounced allosteric effect that affects the loops that regulate access to the catalytic site (44-52 and 185-201 loops). In addition, it affects the residues 137-140, where a specific backbone hydrogen bond is formed between Lys137 and Phe140. The exact role of this conformational change will be the subject of future studies, but the results presented here show that it acts as a relay between domain III and the catalytic site.

Given that conformational changes in the catalytic site are observed when Ebselen binds to Mpro far away from this specific region, we turn our attention to the hydration characteristics of the catalytic site in the Mpro-Ebselen complex (when bound to domain II-III interface) and in the Mpro-apo structure (PDB code: 6m03) (20). AQUADUCT 1.0.5 (21) was used to analyze the water structure and water flux in the Mpro protein, with a time window of 50 ns, and sampling every 1 ps. The results are shown in **Figure 5**.

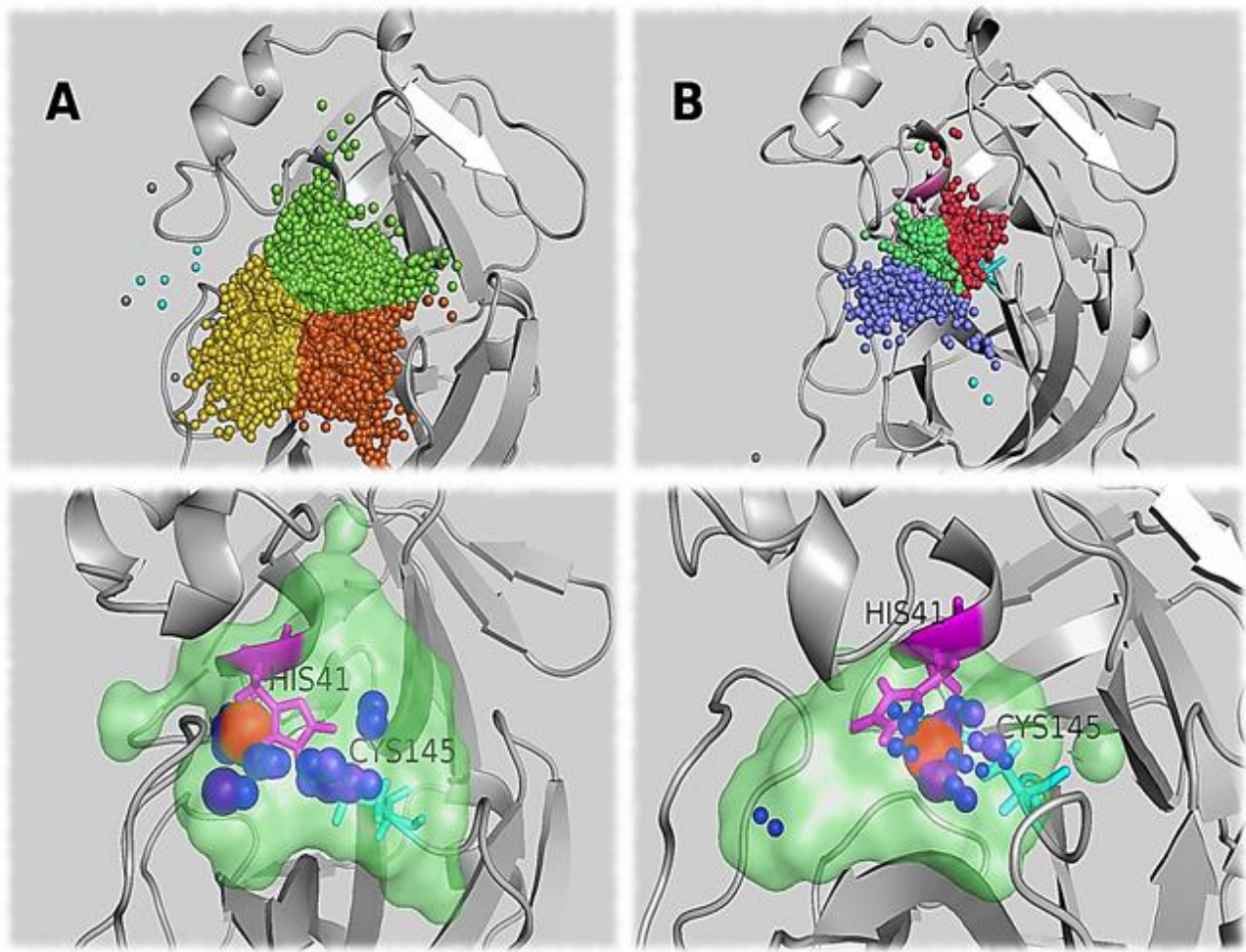

**Figure 5.** Upper: Water inlet clusters for (**A**) apo protein and (**B**) Ebselen in Domain II-III. The biggest cluster in both cases is divided into three parts using K-means (21) and colored for clarity. Lower: Maximum Available Volume (MAV) (as light green surface) and hot-spots (blue-red color and sphere radius relative to occupation values) for (**A**) apo and (**B**) Ebselen in Domain II-III. The catalytic dyad is represented with sticks (magenta: His41, cyan Cys145).

**Figure 5** (upper panel) shows that Ebselen binding between Domain II-III leads to fewer water inlets compared with the Mpro apo state; this is indicative of a water flux reduction in the catalytic site due upon Ebselen binding, even though it happens far from the catalytic site. From the lower panel in **Figure 5**, it is evident that a volume reduction of the catalytic site occurs when Ebselen is located between domain II-III compared with the apo state; the Maximum Available Volume (MAV) for water in the catalytic region is around 50% smaller in this case. Note that in the apo protein there is a catalytic water that forms a catalytic triad together with Cys145 and His41 (16,23), and this catalytic water (red in **Figure 5A** lower panel) is preserved and remains close to His41 in the simulations. In contrast, it is clear from **Figure 5B** (lower panel) that the presence of Ebselen induces a displacement and reconfiguration of water hotspots, including the catalytic water (red). Importantly, these effects could prevent the normal enzymatic function of Mpro, as this catalytic water displacement might damage the catalytic triad that is required for protein activity (24). Similarly, the observed pocket volume reduction might affect the accessibility of the polyprotein that Mpro cleaves, thereby reducing enzymatic function.

## Conclusion and discussion:

As mentioned before, Mpro is an attractive drug target against the COVID-19 virus due to is its central role in the viral life cycle. A previous structural and evolutionary investigation suggested that the SARS-CoV-2 Mpro is not a suitable target for de novo development of inhibitors or the repurposing of drugs against the previous SARS coronavirus (16). That study, however, only compared the active sites in Mpro for COVID-19 and the highly similar previous SARS-CoV Mpro in terms of flexibility and plasticity, where major differences in both shape and size were observed indicating that repurposing SARS drugs for COVID-19 may not be effective. Based on their evolutionary analysis, these authors also pointed out that the virus's mutability will pose further challenges to treatments against the COVID-19 Mpro protein. An alternative to this discouraging scenario, however, would be to target the region between the II and III domains, which is implicated in dimer formation.

Here we find that there are *two*, highly probable interaction sites between SARS-CoV-2 Mpro and Ebselen. One is located within the catalytic cavity and, importantly, a second site is in the region between the II and III domains, which is essential for Mpro dimerization (15). Detailed calculations of the free energy reveal a higher binding affinity of Ebselen to the Domain II-III than to the catalytic site.

The strain analysis reveals that Ebselen bound between the II-III domains exert a pronounced allosteric effect that affects the loops regulating access to the catalytic site. In addition, it also affects residues 137-140, where a specific backbone hydrogen bond is formed between Lys137 and Phe140.

The catalytic site water analysis indicates that the proposed allosteric inhibition by Ebselen could occur through a volume reduction of the catalytic pocket, and a reconfiguration of water hotspots in that region. Given the catalytic role of water in this enzyme's activity, these effects could act to prevent the regular enzymatic function of the SARS-CoV-2 Mpro protein.

The discovery of a distant binding site is encouraging in that it offers potential for Ebselen as a new drug lead for the design and development of new Mpro inhibitors and, eventually, COVID-19 treatment. More generally, this result is helpful in that it emphasizes that distant sites could be effective targets for new drugs. Note that current massive virtual screening campaigns are focused primarily on targeting the SARS-CoV-2 Mpro catalytic site (25). Of course, additional structural characterization experiments are necessary to validate the predictions presented in this work, particularly those pertaining to the distant binding site.

A previous evolutionary study (16) showed that the mutation of a few residues belonging to the catalytic site is energetically unfavorable. Therefore, residues such as P39, R40, P52, G143, G146, or L167 could be considered as key anchoring residues for Mpro inhibitor design. The insights put forth in this work have the potential to facilitate the rational molecular design of new analogs, based on the Ebselen scaffold, that result in anchoring at those positions. This alternative approach, uncovered through the extensive molecular simulations presented here, highlights the need to develop reliable, high-throughput methods to screen drug-protein interactions at the molecular level and that incorporate the role of explicit water morelcules. We conclude by pointing out that our focus in this work has been on non-covalent complexes between Ebselen-Mpro. In a subsequent stage, we also plan to investigate covalent complexes involving Cys145.

**Materials and Methods:**

A total of more than 6 μs of classical MD simulations of SARS-CoV-2 Mpro apo state and Mpro-Ebselen complex were run using AMBER18 (28) simulation package (3 μs: ebselen as a molecular probe; 2.4 μs: shear strain analysis; 100ns: water structure and flux analysis; 990ns: free energy analysis). The receptor initial configuration for the Mpro-Ebselen system was taken from recently reported structure for Mpro-N3 inhibitor (5) (PDB ID: 6lu7); where the inhibitor and crystallographic water molecules were removed before starting simulations. Force field parameters for Ebselen were determined using the Antechamber program and described by the General Amber Force Field (GAFF) (26, 27). The partial atomic charges were determined by the restrained electrostatic potential (RESP) fitting technique. Those electrostatic potential calculations were performed at the HF/6-31G level with Gaussian 09. For Mpro apo simulations we used the recently reported structure (20), (PDB ID: 6m03). Approximately 20,000 TIP3P water model molecules and 4 Na$^+$ ions were added. All simulations were carried out using the ff14SB force field (29). The simulation protocol included a first minimization of 7000 steps, involving 3500 steepest descent steps followed by 3500 steps of conjugate gradient energy minimization, where constraints were applied on the protein heavy atoms (force constant 500 kcal x mol$^{-1}$ x Å$^2$) and a second minimization (7000 steps) with no constraints of conjugate gradient energy minimization. Next, during the first equilibration, the temperature was gradually increased from 0 K to 300 K over 50 ps using a Langevin thermostat with a temperature coupling constant of 1.0 ps in the canonical ensemble. Density equilibration and production runs were carried out using a constant pressure ensemble (NPT). All simulations were performed using periodic boundary conditions and a 2 fs time step. Long-range electrostatic interactions were calculated using the Particle Mesh Ewald method with a non-bonded cut-off of 10 Å and the SHAKE algorithm was used to implement rigid constraints.

### Ebselen Global Density map:

In order to determine the most probable interaction sites between Mpro-Ebselen (global hot-spot), 15 replicas of 200 ns each were run (a total of 3 μs), where configurations were saved every 20 ps. The CPPTRAJ (30) software was used to process the trajectories, where at the first stage all trajectories were align by means of minimization the distance among protein backbone atoms (C, N, CA). Grid command was used to track Ebselen molecule and to produce number density map, where the grid resolution was selected to be 0.5 Å.

### Strain analysis:

Initial Mpro-Ebselen complex configurations were selected from previous global binding affinity study, with Ebselen being bound to the catalytic site, to the intersection between Domain II-III, and to both of these sites simultaneously. Three replicas of 200 ns for each of these initial setups were run. As a reference simulation, we used the Mpro-apo structure (PDB code: 6m03), and 3 replicas of 200 ns each were run. In this way, a total of 2400 ns of classical MD were run.

To apply the strain formalism from continuum theory to discrete, atomistic systems, differential operators replace the derivatives (17,18,19,31,32). A radius $R$ around each central atom $i$ containing $n$ other atoms $j$ defines the local neighborhood around the central atom. The instantaneous position of atom $i$ at any timestep in the MD is $x_i$, and the position of the same atom at any timestep of the reference simulation is $x_{0,i}$. To first order, the distances between atom $i$ and its neighbors $j$ are related through the deformation matrix **F** by $x_j - x_i = F(x_{0,j} - x_{0,i})$, which forms an overdetermined system of linear equations, and an optimized $F^*$ is sought by minimizing the difference between the actual distances and the projected distances to an affine deformation: $min \sum_{j=1}^{n} [x_j - x_i - F(x_{0,j} - x_{0,i})]^2$. The atomic strain tensor is then found by $\varepsilon = \frac{1}{2}(F^T F - I)$, and the magnitude is defined as the L2 norm of the shear part of the strain tensor: $\| \left( \| - \frac{1}{3} \| \| \cdot I \right)$, since proteins are generally incompressible (17,19). For this analysis, a radius of 10 Å around each atom is considered (19), and only the Cα atoms are used in the calculations. The reference simulation is the apo protein trajectory and the strain is then measured using the Ebselen simulations to elucidate the effect of binding of Ebselen at the different sites.

In addition, β-factors were estimated over the same trajectories using the atomicfluct command of the CPPTRAJ module of Amber18.

### Aqua Duct:

To analyze the water structure and water flux in the Mpro protein of SARS-CoV-2 we used AQUADUCT 1.0.5 (21). We obtained Inlets, Maximum Available Volume (MAV) and Hotspots for water molecules in the Mpro catalytic region. For calculations, this region was defined as a 5 Å sphere around the center of geometry of the active site residues (H41, C145, H164, D187) (16). The Mpro protein was studied in two different scenarios, with Ebselen in the Domain II-III site and the apo protein with no ligand. The time window used in both calculations was 50ns, sampling every 1 ps. Images were created with open-source PyMOL (22).

### Free Energy:

The absolute binding free energy is defined as: $\Delta G_{binding} = \Delta G^L - \Delta G^{RL}$, where $\Delta G^{RL}$ is the free energy change of Ebselen annihilation in the Mpro complex, and $\Delta G^L$ is the free energy change of Ebselen annihilation in water. To calculate these free energy changes, we use Thermodynamic Integration (TI) implemented in PMEMD for Amber 18. We use the one step anhihilation protocol with soft core potentials (34). In addition, we adopted a simple approach with multiple runs starting from the most probably binding cluster estimated from previous MD simulations. In this way, three independent replicas for each site were taken into account, as well as three replicas for Ebselen solvated in pure water. Eleven equally spaced windows were used (ΔLambda=0.1) with 10ns of simulation time per window. To keep the ligand from wandering in TI calculations, we used a soft restraint of 10 kcal/molÅ$^2$ (35).